\documentclass[aip,cha,twocolumn,groupedaddress,showpacs,showkeys,preprintnumbers,amsmath,amssymb]{revtex4-1}

\usepackage{dcolumn}
\usepackage{bm}
\usepackage{graphicx}    
\usepackage{times}
\usepackage{hyperref}
\usepackage{color}

\begin{document}

\title{Detecting Functional Communities in Complex Networks}

\author{Sanjeev Chauhan}
\author{Michelle Girvan}
\author{Edward Ott}
\affiliation{Institute for Research in Electronics and Applied Physics,\\ University of Maryland, College Park, Maryland 20742, USA}

\date{\today}

\begin{abstract}
We consider an alternate definition of community structure that is functionally motivated.  We define network community structure-based on the function the network system is intended to perform.  In particular, as a specific example of this approach, we consider communities whose function is enhanced by the ability to synchronize and/or by resilience to node failures.  Previous work has shown that, in many cases, the largest eigenvalue of the network's adjacency matrix controls the onset of both synchronization and percolation processes.  Thus, for networks whose functional performance is dependent on these processes, we propose a method that divides a given network into communities based on maximizing a function of the largest eigenvalues of the adjacency matrices of the resulting communities.  We also explore the differences between the partitions obtained by our method and the modularity approach (which is based solely on consideration of network structure).  We do this for several different classes of networks.  We find that, in many cases, modularity-based partitions do almost as well as our function-based method in finding functional communities, even though modularity does not specifically incorporate consideration of function.
\end{abstract}


\maketitle

{\bfseries
Many methods for finding community structure in complex networks are based on structural criteria.  Some of these methods maximize the number of links with in communities, while others consider the pattern of interconnection between the nodes, etc.  Here, we consider situations in which network communities are thought to have functional purposes, and we consider the implications of the hypothesis that the communities form to enhance their function.  As an example, we specifically consider communities that are thought to form such that they have better synchronizability and/or greater tolerance to random node deletions.  We propose a method to identify communities that have these enhanced dynamical properties, by maximizing a spectral function.  Since the network function-based features may or may not be captured well by the structure-based methods, we also explore the differences between the method proposed in this paper and the widely used structure-based modularity method.  Somewhat surprisingly, we find that, in many cases, there is good agreement between our function-based community partition, and modularity based partitions.  This suggests enhanced utility of the structure-based method, even when functionality is strongly implicated in community formation.
}

\section{Introduction}
\label{introduction}

Community structure has been shown to exist in many social, biological and technological networks \cite{girvan2002,palla2007,barabasi2004,adamic2005,gleiser2003}.
Intuitively, we understand a community as a group of network nodes that ``interact'' more strongly with each other than with nodes outside their community.
Community structures can have significant influence on the organization and dynamics of the network.  For example, communities might be substructures that represent functional units, as in some biological systems \cite{spirin2003}.
In this paper, we consider a dynamical definition of communities in which we assume that the communities form such that they have a functional (and hence dynamical) meaning.  Our problem is to identify and characterize communities based on a functional criterion.

Much of the research related to community structure in networks has been directed toward finding the ``best'' possible community partition of a network.  Direct application of traditional computer science and sociological approaches for finding community structure in complex networks has been shown to be problematic \cite{girvan2002,newman2004,newman2006}.  Various methods have been proposed for detecting community structure in complex networks, \emph{e.g.}, the edge betweenness method \cite{girvan2002}, the eigenvector method \cite{newman2006}, methods based on simulated annealing \cite{guimera2005}, synchronization dynamics \cite{arenas2006,boccaletti2007}, $k$-clique percolation \cite{palla2005}, link communities \cite{ahn2010}, etc.  Many community finding methods are based on modularity \cite{modularity2004}, which, for a given partition of nodes into communities, gives a structural measure of the goodness of that partition.  In the definition of modularity, a community is considered to be a group of nodes within which connections are relatively dense compared to a suitable expectation.  Reviews of structural based methods for dividing networks into communities (with most based on modularity) can be found in Refs. \cite{newman2004,danon2005}.  An excellent overall review on community structure can be found in Ref.\cite{fortunato2010}.

Our motivation for this paper is that, as discussed above, in the past, the definition of a community has often been based on the structural features of networks, \emph{e.g.}, modularity.  In this paper, we will adopt the view that, in many situations, the most appropriate way of defining a community may depend on the application that the resulting division will be used for, which, in turn, depends on the \emph{function} of the network.
For instance, we may desire a different definition of community structure if we are trying to find clusters of friends in a social network than if we are trying to find metabolic pathways in a biochemical network.
One expects that a method designed for a particular consideration may not necessarily work in other situations.
In this paper, as an example, we consider a particular network function and propose an alternate definition of communities for this kind of function.
Specifically, we consider communities that are thought to form so that they have better synchronizability and/or robustness to random node failures.
Our method is based on the observation that a network's function is enhanced when the maximum eigenvalue, $\lambda_*$, of their adjacency matrix is large.  Examples where this applies include synchronization of network coupled phase oscillators \cite{restrepo_dir2006,restrepo2005} and percolation on directed networks \cite{restrepo2008}.
Although we specifically consider directed networks in this paper, the method can also be used to find communities in undirected networks.

It is not obvious that the partitions obtained using a structure-based method will also correspond to good functional partitions.  To analyze this, we explored the difference between the eigenvalue maximization method presented in this paper and the widely used modularity method, which is based purely on consideration of network structure.  Although, we find cases where the two methods yield significantly different results (Sec. \ref{biased_networks}), we also find that, in many situations (Sec. \ref{structural_identification}), the partitions that maximize modularity also tend to score highly according to our functional measure, which we found rather surprising.
Thus, our results suggest that, in many cases, modularity maximization is effective in identifying functional communities.

The organization of this paper is as follows.  In Sec. \ref{fn_eig}, we review the largest eigenvalue of the adjacency matrix of networks without community structure and its relation to network functional properties.  In Sec. \ref{define_fn}, we define a largest-eigenvalue-based measure that can be used to determine community structure in networks.  In Sec. \ref{methods}, we describe the method used to detect community structure given our functional definition.  The construction of networks with eigenvalue based communities is also discussed. In Sec. \ref{results}, we give results for the method proposed in this paper and compare these results with results from the modularity approach.

\section{Network functions and the largest eigenvalue of the adjacency matrix}
\label{fn_eig}

The largest eigenvalue of a network's adjacency matrix in the absence of community structure can be used to characterize both synchronization and percolation phenomenon.  In this section, as background, we discuss the significance of the largest eigenvalue of network adjacency matrix for these network functions.

\subsection{Synchronization}
\label{lareig_synch}

Synchronization is a population effect that emerges in many complex systems composed of a large number of dynamical components \cite{strogatz2000}.  The classical model of Kuromoto describes the synchronization of phase oscillators that are uniformly globally coupled and have natural frequencies drawn from a heterogeneous distribution \cite{kuramoto1984}.  In the limit of large network size, a phase transition, separating the synchronized and the unsynchronized states, is observed for the Kuramoto model.  For synchronization on networks with large average degree and arbitrary degree distribution, similar results have been reported \cite{restrepo_dir2006,restrepo2005}.

For synchronization of phase oscillators in complex networks, the evolution,
\begin{equation} \label{eq:syn_eq}
\dot{\theta}_i= \omega_i + K \sum\limits_{j=1}^{N} A_{ij}~\sin(\theta_j-\theta_i),
\end{equation}
is considered, where $\theta_i$ and $\omega_i$ are the phase and intrinsic frequency of the $i^{th}$ oscillator, $K$ is an overall coupling strength, and $N$ is the number of nodes in the network.  Here, $A_{ij}$ is the $(i,j)^{th}$ entry of the adjacency matrix which has value $1$ if there is a link from node $j$ to node $i$; otherwise it is $0$.  The synchronization of nodes in the network can be characterized by the global order parameter, $r$, given by
\begin{equation} \label{eq:order_parameter1}
r = \left| \frac {\sum_{j=1}^N~e^{i\theta_j}} {N} \right|.
\end{equation}
Perfect synchronization (typically occurring for $K\rightarrow\infty$) corresponds to $r=1$.  For large $N$, synchronized and unsynchronized behaviors of the system are signified by a value of $r$ significantly above zero and close to zero, respectively.

For networks with large average degree and an arbitrary degree distribution, results based on mean field theory show that the critical value of coupling strength, which separates the synchronized and unsynchronized states, is determined by the first two moments of the degree distribution of the nodes \cite{ichinomiya2004,lee2005}.  Restrepo \emph{et al.} obtained better estimates of the critical coupling strength in the case of directed networks \cite{restrepo_dir2006}.  In particular, they show that the critical value of the coupling strength, $K_c$, is determined by the largest eigenvalue of the network adjacency matrix,
\begin{equation} \label{eq:k_c}
K_c=\frac {K_0} {\lambda_*},
\end{equation}
where $K_0$ is a constant which depends on the distribution of oscillator frequencies and is independent of the network characteristics.  Thus, the higher the largest eigenvalue of the network adjacency matrix, the smaller the value of $K$ needed to attain the phase transition to synchronization for such networks.

\subsection{Percolation}
\label{lareig_perco}

Percolation is another network property that has been studied extensively.  In the percolation model, a phase transition separates two phases characterized by the presence and absence of a giant connected component when nodes (site percolation) or links (bond percolation) are removed from the network.
In undirected networks that do not have any degree correlations between linked nodes, the percolation transition has been shown to depend on the second moment of the degree distribution \cite{cohen2000}.
Percolation in case of directed networks has also been explored (\emph{e.g.} see Refs. \cite{newman2001,dorogovtsev2001,schwartz2002,boguna2005}).

Some approaches focus on a Markovian approach for studying percolation phenomenon \cite{schwartz2002,boguna2005,vazquez2003,serrano2006}.  Restrepo \emph{et al.} \cite{restrepo2008} studied the percolation problem without the need of a Markov network model but requiring the knowledge of the network adjacency matrix.  For directed networks that are locally tree like, they found that the percolation transition occurs when a fraction of nodes,
\begin{equation} \label{eq:percol_tran}
p_c=1 - \frac {1} {\lambda_*},
\end{equation}
have been randomly removed from the network.  This indicates that when the largest eigenvalue, $\lambda_*$, of the network adjacency matrix is high, the network can tolerate a large number of node deletions before it disintegrates.

\section{A functional definition of community structure using eigenvalues}
\label{define_fn}

As discussed in Sec. \ref{fn_eig}, in the case of directed networks without community structure, larger values of $\lambda_*$ make the network more resilient to breaking up into many disconnected pieces when nodes are randomly removed (e.g., due to failure or attack).  Furthermore, synchronization in a heterogeneous collection of phase oscillators is promoted by increasing $\lambda_*$.  This suggests that, if a network's function depends on synchronization of heterogeneous oscillators and/or robustly maintaining connectivity, then consideration of the largest eigenvalues of the adjacency matrices of individual communities may provide a natural basis for a useful functional definition of community structure on such networks.

We propose a measure that is meant to quantify the strength of network division into communities that have better synchronizability and robustness to random node failures.  Motivated by the role of the largest eigenvalue in both synchronization and percolation, our measure sums a monotonically increasing function of the largest eigenvalues of the communities.  We view this as an example of a functional definition of communities that might be appropriate in some cases, but we also emphasise that other definitions would be appropriate for other purposes.

\begin{figure}[tb]
\centering
\includegraphics[width=5.6cm,height=4.5cm,angle=0]{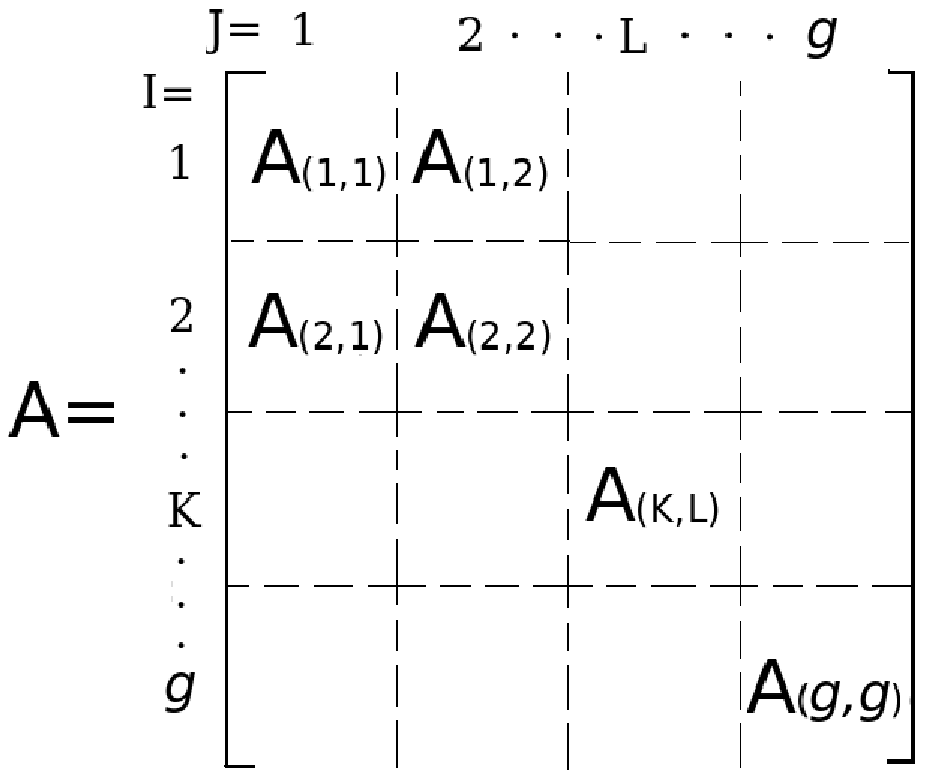}
\caption{Adjacency matrix of a network with $g$ communities, in block matrix form.  Each diagonal block corresponds to the adjacency matrix of a community, while the off diagonal blocks correspond to links between communities.}\label{fig:block_matrix}
\end{figure}

For clarity, we can write the adjacency matrix, $A$, of networks with community structure in block matrix form as shown in Fig.\ref{fig:block_matrix}.  Each diagonal block of $A$ then corresponds to the adjacency matrix of an individual community, while the off diagonal blocks correspond to the links between communities.  We propose that, given a network, if we can find a partition of the network into communities that have higher largest eigenvalues of their corresponding diagonal block adjacency matrices, then those communities will have enhanced network functions.

Specifically, the definition of community structure that we study is as follows:
\begin{enumerate}
\item Consider a partition of a network into $g$ communities.
\item Calculate the maximum eigenvalues ($\lambda_{*1}, \lambda_{*2}, ..., \lambda_{*g}$) of the adjacency matrices of all the communities.  Here, $\lambda_{*k}$ is the largest eigenvalue of the $k^{th}$ diagonal block in Fig.\ref{fig:block_matrix}.
\item Define the ``spectral cohesion'':
\begin{equation} \label{eq:spectral_cohesion}
\Lambda=\sum\limits_{k=1}^g \ln(\lambda_{*k}).
\end{equation}
\end{enumerate}
The spectral cohesion, $\Lambda$, provides a functionally based measure of the community strength of a particular partitioning of the network. We can thus define the \emph{best} division into $g$ communities as the one that maximizes $\Lambda$, where we think of \emph{best} as being with respect to the enhancement of synchronization or resilience.  Note that the definition of communities according to Eq.(\ref{eq:spectral_cohesion}) can be used for both symmetric and asymmetric matrices.  In Sec. \ref{results}, we will demonstrate the utility of this definition for directed networks, in particular.

As an aside, we emphasize that our choice of the spectral cohesion function in Eq.(\ref{eq:spectral_cohesion}) is somewhat arbitrary; \emph{e.g.}, $\Lambda=\sum f(\lambda_{*k})$ for any function $f(\lambda)$ that is monotonically increasing with $\lambda$ might alternatively be considered.  However, we shall, in all of what follows, use $f(\lambda)=\ln(\lambda)$.  This is partly motivated by the analogy to entropy, and by our studies with $f(\lambda)=\lambda^{\beta}$, for $\beta=1$ and 2, which, for several test networks, yielded results that were very similar to those for $f(\lambda)=\ln(\lambda)$.

\begin{figure}[tb]
\centering
\includegraphics[width=4.0cm,height=6.0cm,angle=-90]{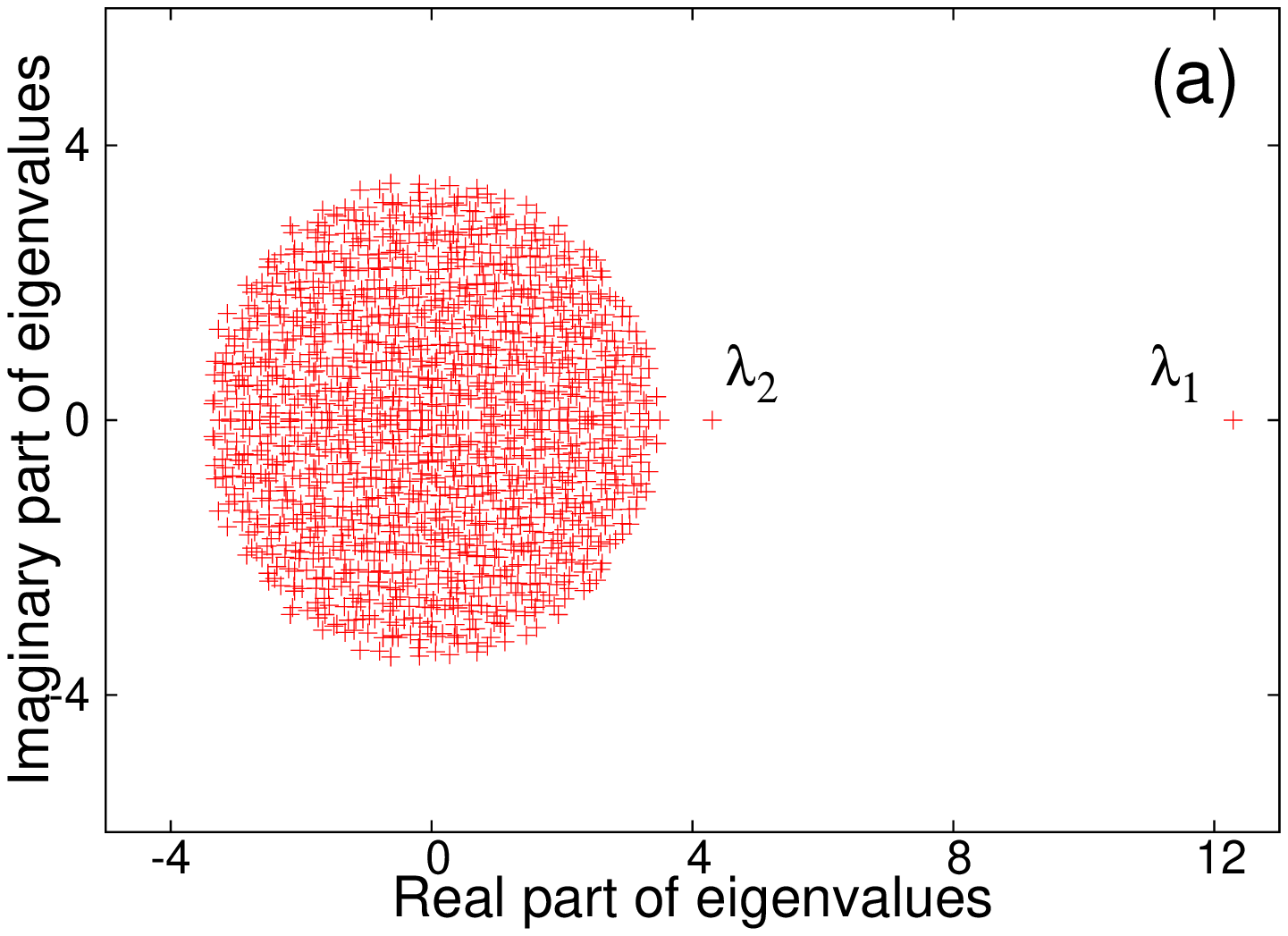}
\includegraphics[width=4.0cm,height=6.0cm,angle=-90]{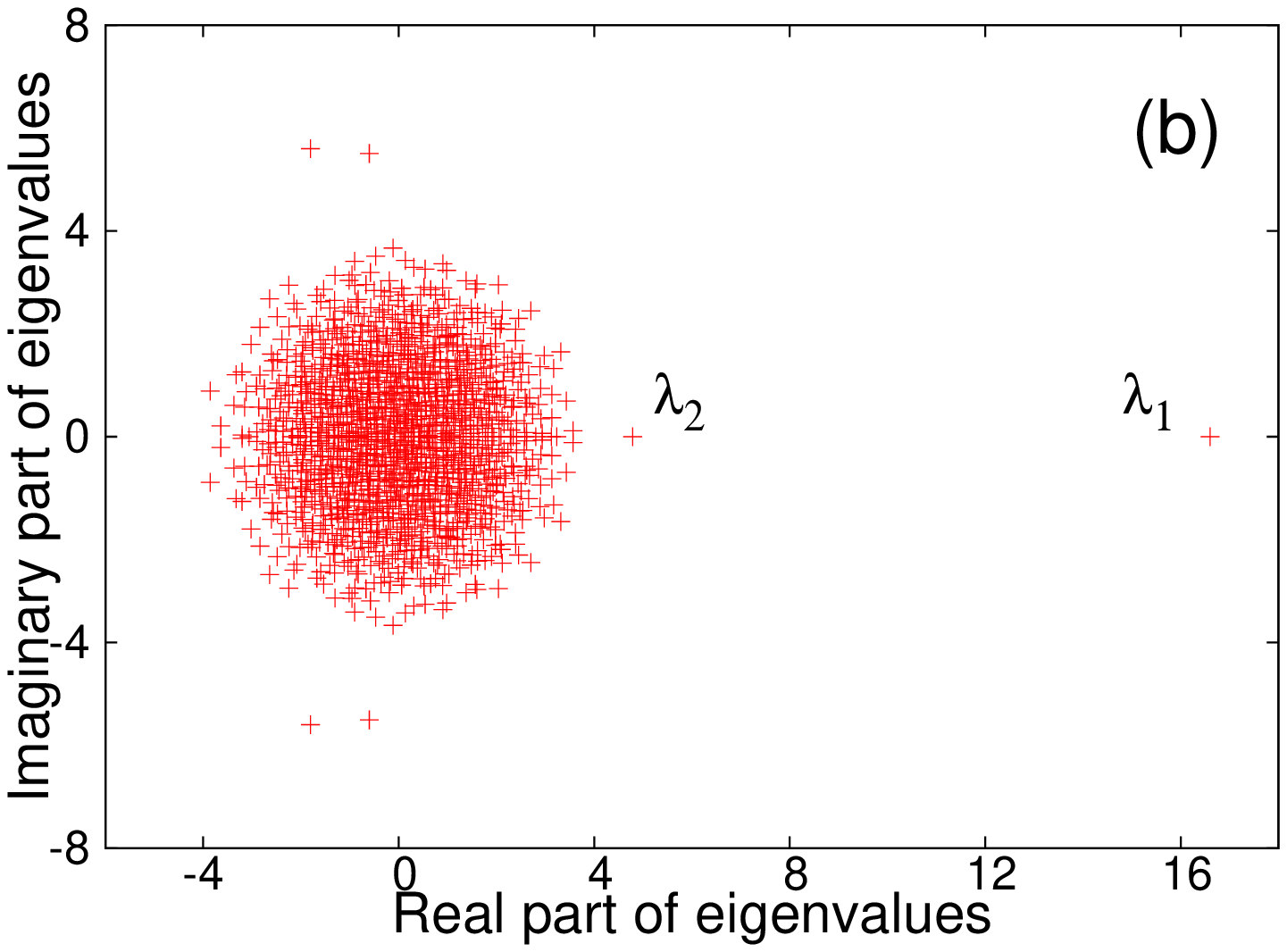}
\caption{(Color online) Eigenvalue plots for (a) a directed Erdos-Renyi type network , and (b) a directed scale-free network with two equally sized communities.  By construction, the nodes in the network have $\langle d \rangle_I = \langle d \rangle_X = 6$, but the communities are defined such that they have maximal directional degree assortativity within them.  The networks have $N=1400$ and $\langle d \rangle = 12$.  Here, $\lambda_1$ and $\lambda_2$ are the largest and the second largest real positive eigenvalues of the network adjacency matrix.}\label{fig:eigenvalPlots_assort}
\end{figure}

While the method of maximizing the spectral cohesion, $\Lambda$, gives us the best division into $g$ communities, it does not tell us how to choose the appropriate value of $g$, i.e., the number of communities that the network contains.  In an another paper \cite{paper1}, we showed that the number of communities in the network may be obtained from the eigenspectra of the adjacency matrix of the full network.  Specifically, we showed that for networks with communities, there typically exists a relatively small set of $g^*$ positive real adjacency matrix eigenvalues that are significantly larger than, and well separated from the large number of other eigenvalues.  When a network has strong functional communities, we expect them to have higher values of their largest eigenvalues.  The number of $g^*$ large positive eigenvalues that are well separated from the rest of the other eigenvalues, can be taken as an appropriate choice for $g$.  Reference \cite{paper1} provides examples of eigenvalue plots for networks with community structure that have high density of links within communities and lower density of links between communities.  In Fig. \ref{fig:eigenvalPlots_assort}, we give examples of the eigenvalue plots for other types of networks with community structure.  Figure \ref{fig:eigenvalPlots_assort}(a) is for an Erdos-Renyi type directed network, and Fig. \ref{fig:eigenvalPlots_assort}(b) is for a scale-free directed network.  Both networks have two communities of equal sizes, $N=1400$ and $\langle d \rangle = 12$, where $\langle d \rangle$ for a directed network here denotes average in-degree or out-degree, which are both equal.  In order to briefly describe our choice of networks for Fig.\ref{fig:eigenvalPlots_assort}, we first note a result for random networks without community structure.  In particular,
if there is directional degree assortativity (see Eq. (\ref{eq:rho_eq}) for the definition), then a mean field theory (described in Sec. \ref{corr_n_assort}) \cite{restrepo2007} shows that, other things being equal, $\lambda_*$ is larger for networks with larger
directional degree assortativity.  The networks used to generate the plots of Fig. \ref{fig:eigenvalPlots_assort} have the property that the average number of in/out-links that connect a node to nodes in its own community, $\langle d \rangle_I$, is equal to the average number of in/out-links that connect the node to nodes in the other community, $\langle d \rangle_X$, but the communities are defined such that the communities have maximal
directional degree assortativity within them
(see Sec. \ref{corr_n_assort} for details).  Thus, in the absence of
directional degree assortativity, the networks are random networks with no community structure.

\subsection{Cycles in the graph and the largest eigenvalue}
\label{lambda_and_cycles}

In this paper, our aim is to find communities that have enhanced network functions which in turn depends on the largest eigenvalues of the community adjacency matrices.  Although our choice of $\Lambda$ in Eq.(\ref{eq:spectral_cohesion}) to a certain extent was arbitrary, we can consider a useful interpretation of this function.  Eigenvalues of the community adjacency matrices, $A_k$, are related to the cycles in the communities.  The number of cycles of length $n$ inside a community $k$ equals the sum of the diagonal components of $A_k^n$ which in turn equals $\sum\limits_{i=1}^{N_k} \lambda_{ik}^n$, where $\lambda_{ik}$ is the $i^{th}$ eigenvalue of community $k$.  Thus, for large $n$, the exponential growth of the number of cycles with their length is 
\begin{equation} \label{eq:lambda_n_cycles}
    \lim_{n \rightarrow \infty} \frac{ \ln~ ( \sum\limits_{i=1}^{N_k} \lambda_{ik}^n ) }{n} = \ln~\lambda_{*k},
\end{equation}
when the limit exist.  Thus, when we use $\Lambda$, we expect to find communities that have high number of cycles within them.

\section{Methods}
\label{methods}

\subsection{Detecting functional communities}
\label{annealing}

Thus far, we have proposed a quantity whose maximization, we hypothesize, should yield a good division of a network into communities for the network functions we are interested in.  In this section, we provide an outline for a simulated annealing scheme \cite{kirkpatrick1983} that finds a desirable division of the network.  The advantage of this simulated annealing method is that it can provide a network division whose spectral cohesion is very close to the true maximal value.  The disadvantage is that it is computationally quite intensive.  In order to fairly compare our results with the modularity approach, we also use simulated annealing to find a network division that maximizes the modularity function for a fixed number of communities.  The modularity function is based on a comparison between the number of links connecting nodes in the same community to the number expected in a random network without community structure.  For directed networks, the modularity ($Q$) is defined as \cite{arenas2007,leicht2008}
\begin{equation} \label{eq:modularity}
Q=\frac {1}{m} \sum\limits_{i,j} \left[ A_{ij} - d_i^{in} d_j^{out}/{m} \right] \delta_{c_i,c_j},
\end{equation}
where $d_i^{in}$ denotes the in-degree of node $i$, $d_j^{out}$ denotes the out-degree of node $j$, and $m$ is the number of edges in the network.  $c_i$ and $c_j$ denote the community indices of nodes $i$ and $j$, and $\delta_{c_i,c_j}=1$ $(=0)$ if $c_i=c_j$ $(c_i\neq c_j)$.

In our simulated annealing scheme, we begin by assigning nodes randomly to $g^*$ different communities, where we find $g^*$ as described in Sec. \ref{define_fn} and Ref.\cite{paper1}.  We then choose a node at random and pick a random community, to which to consider moving it.  If this move would result in an increase in the value of the function we are trying to optimize, say $F$ (which could be either $\Lambda$ or $Q$)
, we perform the move.  If the move would result in a decrease in the value of the function, we perform it with Boltzmann acceptance probability $e^{\Delta F/T}$, where $\Delta F < 0$ is the change in the function $F$ and $T$ is the `temperature' (this is the basic Metropolis algorithm \cite{metropolis1953}).  For each temperature value, we repeat this process $\alpha N^2$ times, where $N$ is the number of nodes in the network and $\alpha$ is a chosen factor $\le 1$.  After $\alpha N^2$ iterations, we reduce the temperature by a factor of $0.99$.
Using the parameters described, the whole process is repeated until an asymptotic value of $F$ is reached.

In the case of the spectral cohesion, there is a caveat to the movement of nodes.  In some networks, there are nodes that do not affect the eigenvalue of any of the communities.  If such a node is chosen at a given iteration, it is moved to a community that is randomly selected at that iteration if it has more links to that community than its own, without regard to the directionality of the links.  If it has fewer links to the randomly chosen community, the move is accepted with a probability which depends on the number of links the node has to both the communities.  We expect this strategy to be reasonable only if, as in the numerical examples we treated, we have a small number of such nodes in the network.

Since we are interested only in the largest eigenvalues of the matrices, we use the power method \cite{hoffman2001} to calculate these eigenvalues.
 Assuming that $\lambda_{*k}$ is well separated from the other eigenvalues, for a dense $N_k \times N_k$ matrix, where $N_k$ is the number of nodes in community $k$, the needed computational time for this approach to give the dominant eigenvalue of the matrix scales as $O(N_k^2)$.  In the case of sparse matrices, the required number of operations needed to compute $\lambda_{*k}$ scales as $O(M_k)$, where $M_k$ is the number of non-zero entries in the matrix corresponding to community $k$.  Assuming that we are working with sparse networks with $M_k \sim N_k$ (as is the case with many real-world networks) and that the number of required temperature reductions is independent of $N$ (an optimistic assumption), assuming $N_k \sim N$, this yields an algorithm whose required number of operations scales at best as $O(N^3)$.

When maximizing the spectral cohesion, in many cases, run times of our simulated annealing program can be further reduced by using perturbation theory \cite{restrepo2006} for calculating the estimate of $\Delta \Lambda$ above.  We accept or reject a move based on this estimate of $\Delta \Lambda$.  When a move is accepted, we calculate only the eigenvalues of the communities involved in the change.  When a move is rejected, we go to the next step.  This is much less computationally expensive than recalculating the eigenvalues at each step of the simulated annealing procedure.

The use of perturbation theory for calculating the estimate of $\Delta \Lambda$ is explained as follows.  When a node, say $i$, is chosen at random, we consider moving it from its current community, say $k$, to another community, say $l$, the estimated change in the largest eigenvalue of the adjacency matrix of the node's current community, due to the removal of the node, is calculated using the approximation given in Ref.\cite{restrepo2006}.
Let $V_k$ and $U_k$, respectively, be the left and right eigenvectors of the adjacency matrix of community $k$ corresponding to $\lambda_{*k}$ and normalized so that $V_k^TU_k=1$.  Let $(V_k)_i$ and $(U_k)_i$ denote the components of the vectors $V_k$ and $U_k$ corresponding to node $i$.  Then for $N_k>>1$ and $(V_k)_i (U_k)_i << 1$, removal of node $i$ leads to a change in $\lambda_{*k}$, which is approximately given by \cite{restrepo2006}
\begin{equation} \label{eq:del_f1}
\Delta \lambda_{*k} = - \lambda_{*k} ~ (V_k)_i (U_k)_i.
\end{equation}
Similarly, we estimate the increase in the largest eigenvalue of community $l$, when we consider adding node $i$ to it to be
\begin{equation} \label{eq:del_f2}
\Delta \lambda_{*l} = \frac{(V_l^T \delta A_l)_i(\delta A_l U_l)_i}{\lambda_{*l}}.
\end{equation}
Here, we assumed $\Delta \lambda_{*l} << \lambda_{*l}$ where $\lambda_{*l}$ is the largest eigenvalue of the adjacency matrix of community $l$ before addition of node $i$.  In Eq. (\ref{eq:del_f2}), $\delta A_l$ is the perturbation applied to the adjacency matrix of community $l$ due to the addition of node $i$, $V_l$ and $U_l$ are the left and right eigenvectors of the adjacency matrix of community $l$ corresponding to $\lambda_{*l}$ that satisfy the normalization condition $V_l^T U_l=1$, and $(V_l^T \delta A_l)_i$ and $(\delta A_l U_l)_i$ denote the components of the corresponding vectors corresponding to node $i$.  Note that $\delta A_l$ is of dimension $(N_l+1) \times (N_l+1)$, because when we consider moving node $i$ to community $l$, the number of nodes in community $l$ becomes $N_l+1$.  All the elements of $\delta A_l$ are zero except for the row and column corresponding to node $i$, which has 1's at appropriate locations corresponding to in-links and out-links to and from node $i$ to nodes in community $l$.  The vectors $V_l$ and $U_l$ are $(N_l+1)$ dimensional column vectors, with the entry corresponding to node $i$ being zero.  Thus, for $N_k$, $N_l>>1$, the estimated change in the value of the spectral cohesion $\Lambda$, given by Eq.(\ref{eq:spectral_cohesion}), due to the movement of the node is
\begin{equation} \label{eq:del_f3}
\Delta \Lambda =  \frac {\Delta \lambda_{*k}}{\lambda_{*k}} + \frac{\Delta \lambda_{*l}}{\lambda_{*l}}.
\end{equation}
The above time-saving scheme is especially useful for large networks.  The larger the network, the better the perturbation theory in estimating the change in the spectral cohesion.

Because we use a simulated annealing approach, our method for finding communities is more computationally demanding than many methods that have been proposed that are based on structural definitions of communities.  Our goal here, however, is not to introduce the most efficient algorithm for finding community structure, but rather to test the degree to which a functional approach to finding communities may be appropriate in certain cases.  We leave the development of fast algorithms that identify functional community structure for future work.

\subsection{Construction of test networks with eigenvalue-based communities}
\label{corr_n_assort}

In this section, we give methods for the construction of networks with eigenvalue-based communities.
We will subsequently use these networks for our numerical experiments in Sec. \ref{structural_identification}.
As preliminary preparation for explaining how we construct networks with eigenvalue-based communities, we first note two results relating $\lambda_*$ to the topological properties of networks without communities \cite{restrepo2007}.

\subsubsection{The effect of node in/out-degree correlations}
\label{node_degree_corr}

Considering random directed networks without community structure, if the network is characterized by a joint in/out-degree probability distribution $P(d^{in},d^{out})$, then the expected value of the maximum eigenvalue is \cite{restrepo2007}
\begin{equation} \label{eq:eta_lambda}
\lambda_* = \eta \langle d \rangle,
\end{equation}
where $\langle d \rangle := \langle d^{in} \rangle = \langle d^{out} \rangle$, $\langle ... \rangle$ denotes an average over the network nodes, and $\eta$ is the in/out-degree correlation coefficient,
\begin{equation} \label{eq:eta_eq}
\eta = \langle d^{in}d^{out} \rangle / \langle d \rangle^2.
\end{equation}
Thus, in/out-degree correlation, $\eta>1$ (anticorrelation, $\eta<1$) increases (decreases) $\lambda_*$.  Note that in the absence of node in/out-degree correlation, we have $\eta=1$ and $\lambda_* \approx \langle d \rangle$.

In obtaining the estimate in Eq.(\ref{eq:eta_lambda}), the network is imagined to be constructed by first randomly assigning each node values $(d^{in},d^{out})$ according to $P$, and then randomly linking the nodes accordingly as described for the networks discussed in the Appendix \ref{generating_nets}.

\subsubsection{The effect of directional degree assortativity}
\label{edge_degree_corr}

\begin{figure}[tb]
\centering
\includegraphics[width=3.5cm,height=1.2cm,angle=0]{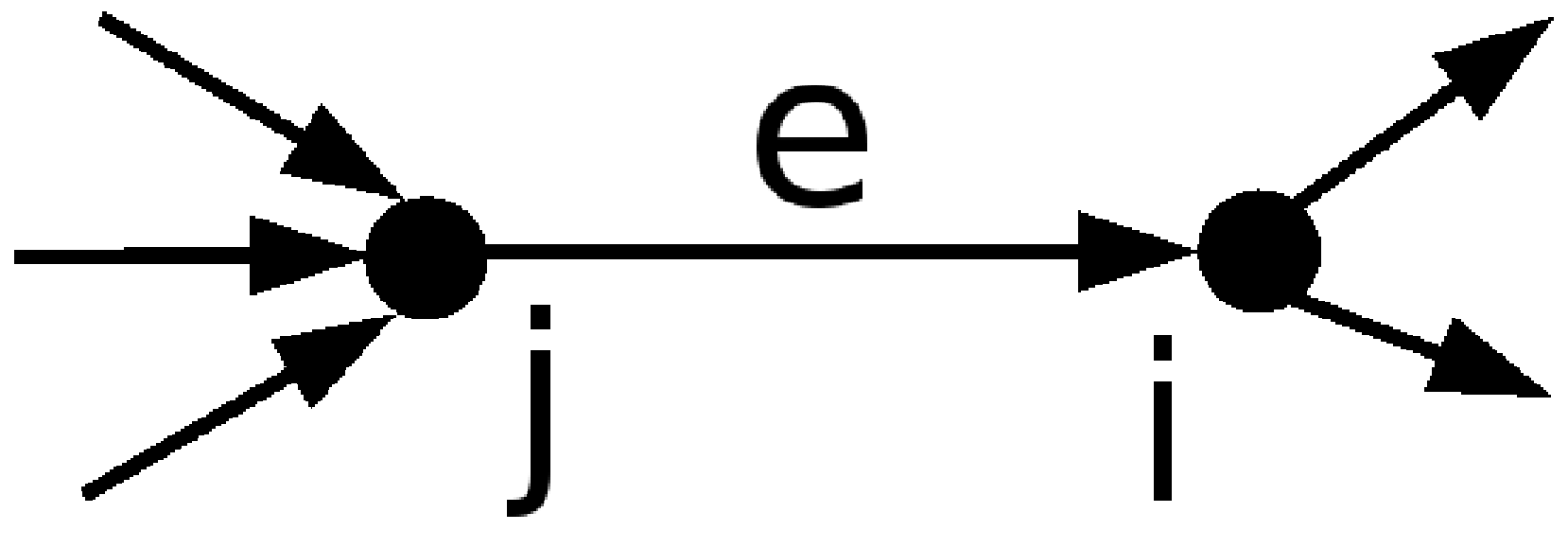}
\caption{An example of one term in the average $\langle ... \rangle_e$ where $d^{in}_j=3$ and $d^{out}_i=2$}\label{fig:one_term_example}
\end{figure}

We now consider random directed networks with uncorrelated in/out node degrees in the distribution $P(d^{in},d^{out})$ that are assortative by degree according to the directed degree assortativity coefficient \cite{restrepo2007},
\begin{equation} \label{eq:rho_eq}
\rho = \langle d_i^{out}d_j^{in} \rangle_e / \langle d_i^{out} \rangle_e \langle d_j^{in} \rangle_e,
\end{equation}
where $\langle ... \rangle_e$ denotes the average over all the edges from node $j$ to node $i$ (Fig.\ref{fig:one_term_example}), but are otherwise random.
In this case, the expected value of $\lambda_*$ is \cite{restrepo2007}
\begin{equation} \label{eq:rho_lambda}
\lambda_* = \rho \langle d \rangle.
\end{equation}
Thus assortativity (corresponding to $\rho>1$) increases $\lambda_*$, while disassortativity ($\rho <1$) reduces $\lambda_*$.

\begin{figure}[tb]
\centering
\includegraphics[width=3.0cm,height=1.5cm,angle=0]{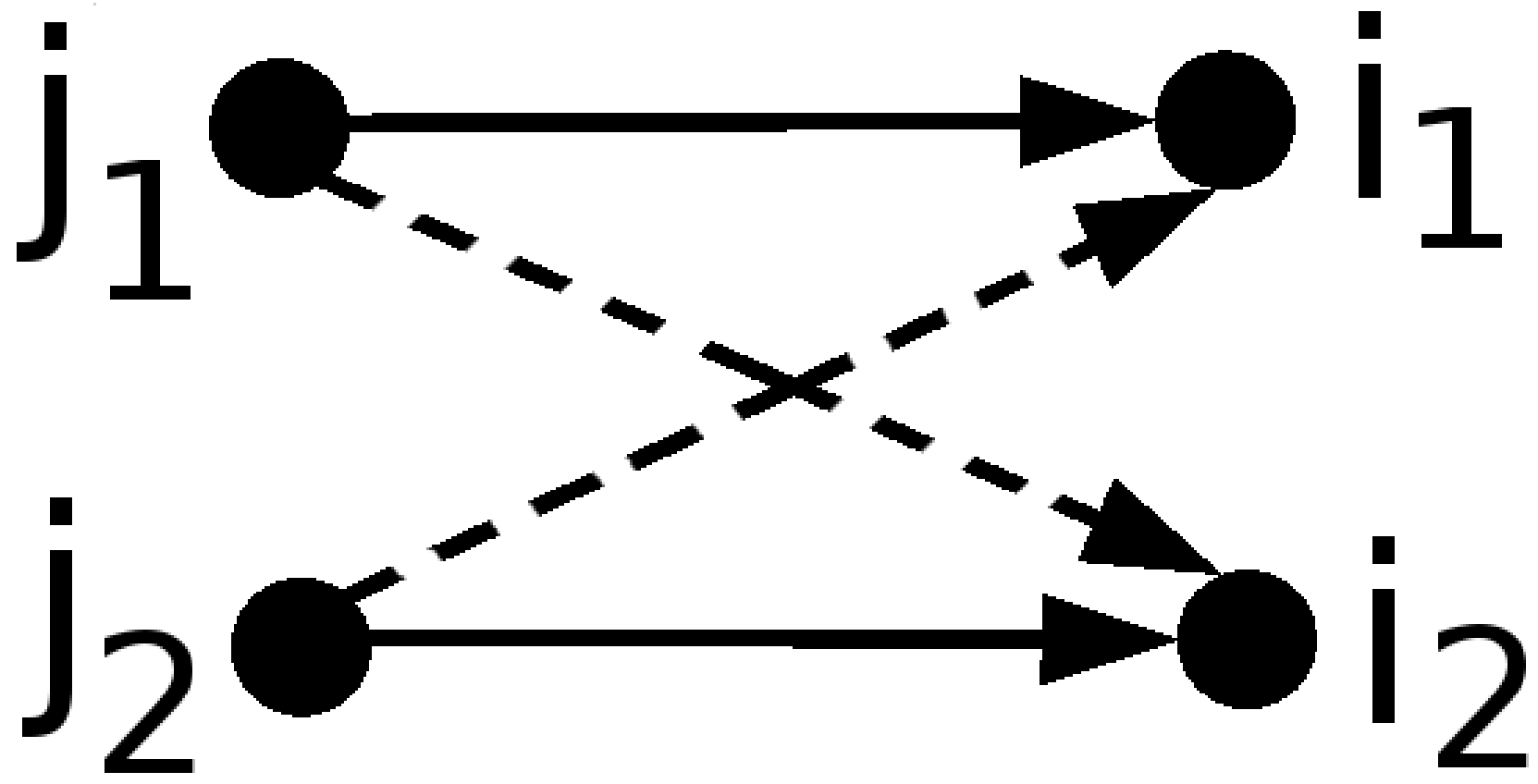}
\caption{Illustration of a destination edge interchange.}\label{fig:edge_interchange}
\end{figure}
Here the network is imagined to be constructed in two stages \cite{restrepo2007}.  First a non-assortative and node degree uncorrelated network is randomly constructed (see Appendix \ref{generating_nets}).  Such a network will have $\rho \approx 1$ for large $N$.  Next, to increase $\rho$ to any desired target value, we first randomly choose two edges, $(j_1 \rightarrow i_1)$ and $(j_2 \rightarrow i_2)$ (see Fig.\ref{fig:edge_interchange}).  We then imagine that we interchange the destinations of these two links, thus producing two new links, $(j_1 \rightarrow i_2)$ and $(j_2 \rightarrow i_1)$.  If $\rho$ increases, we implement the change; if $\rho$ decreases, we do not.  We then randomly choose two new links, and successively repeat this process until $\rho$ approximately reaches its target value.

\emph{Test networks}: The above results can be used as a basis for the construction of networks with eigenvalue-based community structure.  For example, consider networks with two nominally equally sized communities.
The communities can have any ratio of within to between community links but they also have $\eta_c$ or $\rho_c$ greater than one.
Here $\eta_c$ and $\rho_c$ are defined by Eqs. (\ref{eq:eta_eq}) and (\ref{eq:rho_eq}) but with consideration restricted to only those nodes and links that lie within a community under consideration, and thus only using within community node degrees.  That is, we produce higher maximum eigenvalues for the communities by increasing the within community in/out-degree correlation or directional degree assortativity.  We consider both directed scale-free and Erdos-Renyi type networks of these types.  In Sec. \ref{structural_identification}, we will use such networks in numerical experiments.

All the test networks of the type described above that are used in Section \ref{structural_identification}
have $N=1400$ with two nominally equally sized communities.  In our test networks in Section \ref{structural_identification},
we keep $\langle d \rangle_I = 6$ while changing $\langle d \rangle_X$.  By doing this we make sure that the communities have same maximal values of directional degree assortativity and node degree correlations within them as we change $\langle d \rangle_X$.
For the scale-free networks, the maximal attainable $\eta_c$ was approximately 2.12, while the maximal attainable $\rho_c$ was approximately 2.05.  For the Erdos-Renyi type networks, the corresponding maximal attainable values were approximately 1.16 for both $\eta_c$ and $\rho_c$.  In Section \ref{structural_identification},
we used these maximal situations such that both the communities either have maximal $\rho_c$ and $\eta_c \approx 1$, or have maximal $\eta_c$ and $\rho_c \approx 1$.
In these situations, the values of $\lambda_{*k}$ are substantially larger than would be obtained for a random partition of the network into two equally sized communities.  In addition, for the networks with maximal directional degree assortativity within communities, in the eigenvalue plots in Fig. \ref{fig:eigenvalPlots_assort}, we see two positive eigenvalues outside the cloud of the rest of the eigenvalues even with $\langle d \rangle_I = \langle d \rangle_X$, indicating the presence of two communities.
More details on the methods of constructing test networks with eigenvalue based communities are given in the Appendix \ref{generating_nets}.

\section{Results}
\label{results}

In this section, we report results from using the functionally motivated definition of community structure proposed in this paper.  For comparison, we also present results using the modularity method to find partitions in both artificial and real networks.

\subsection{Structural identification}
\label{structural_identification}

\begin{figure*}
    \begin{center}
    \begin{tabular}{ccc}
        \resizebox{55mm}{!}{\includegraphics[width=3.75cm,height=5.00cm,angle=-90]{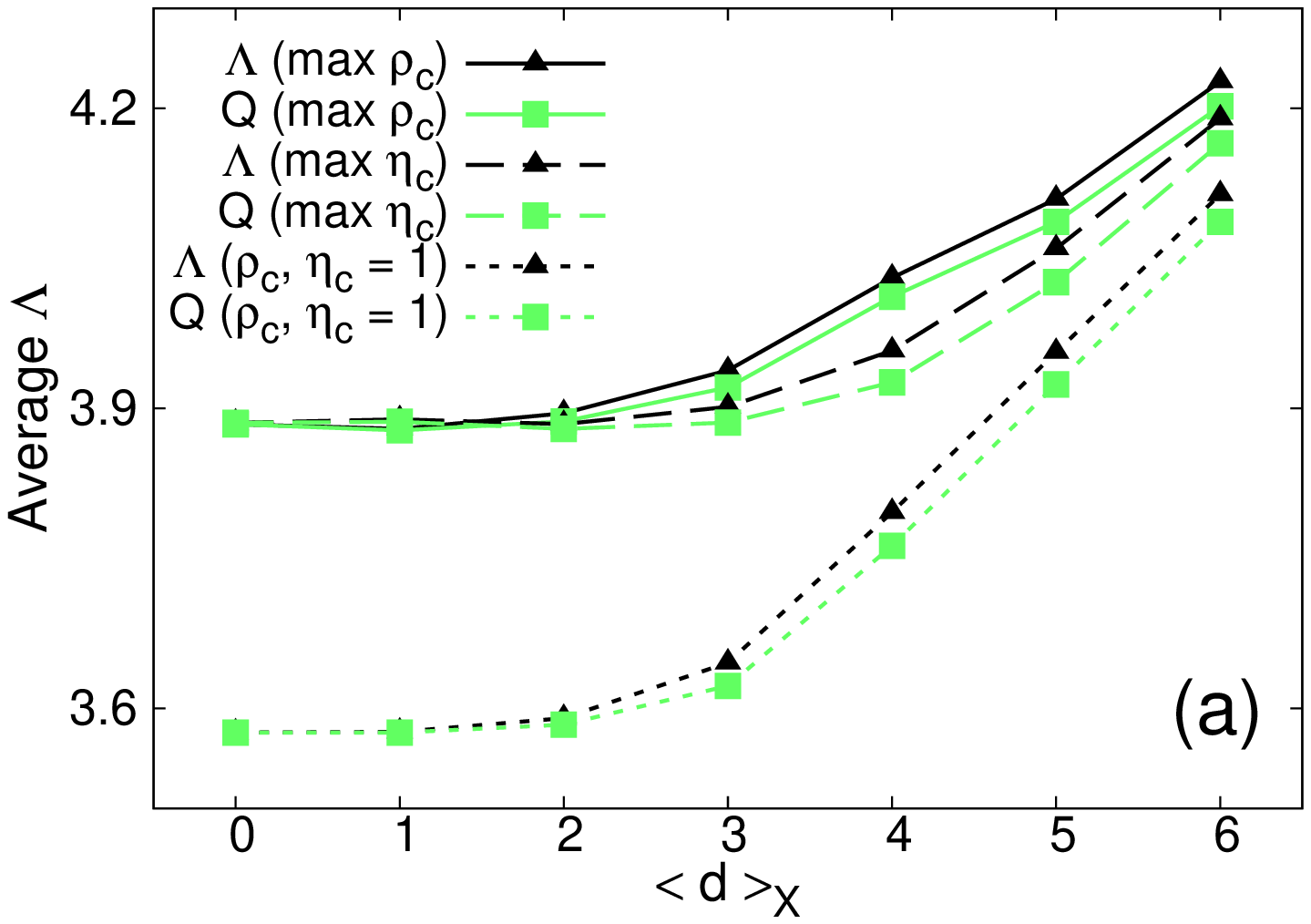}} &
        \resizebox{55mm}{!}{\includegraphics[width=3.75cm,height=5.00cm,angle=-90]{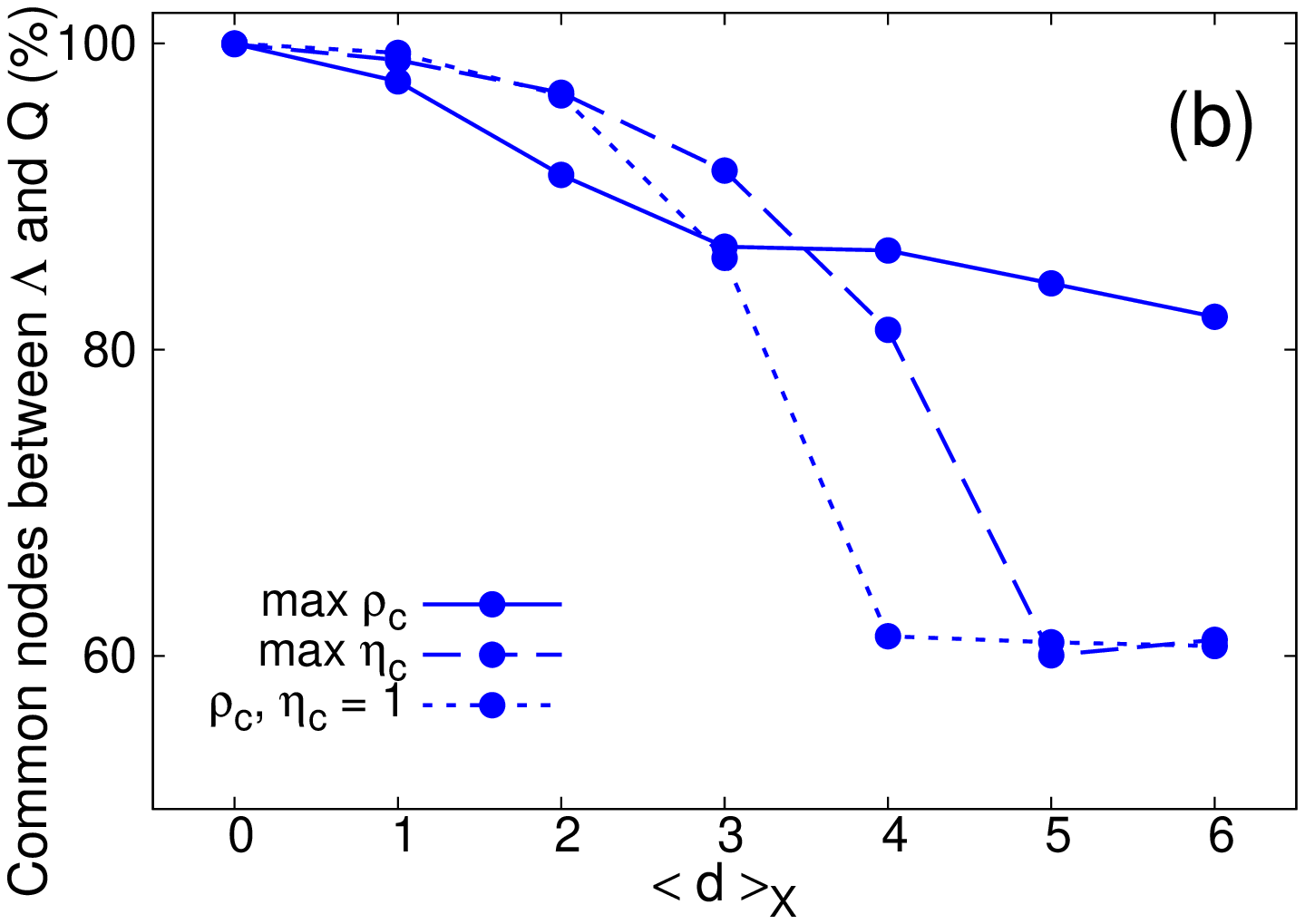}} &
        \resizebox{55mm}{!}{\includegraphics[width=3.75cm,height=5.00cm,angle=-90]{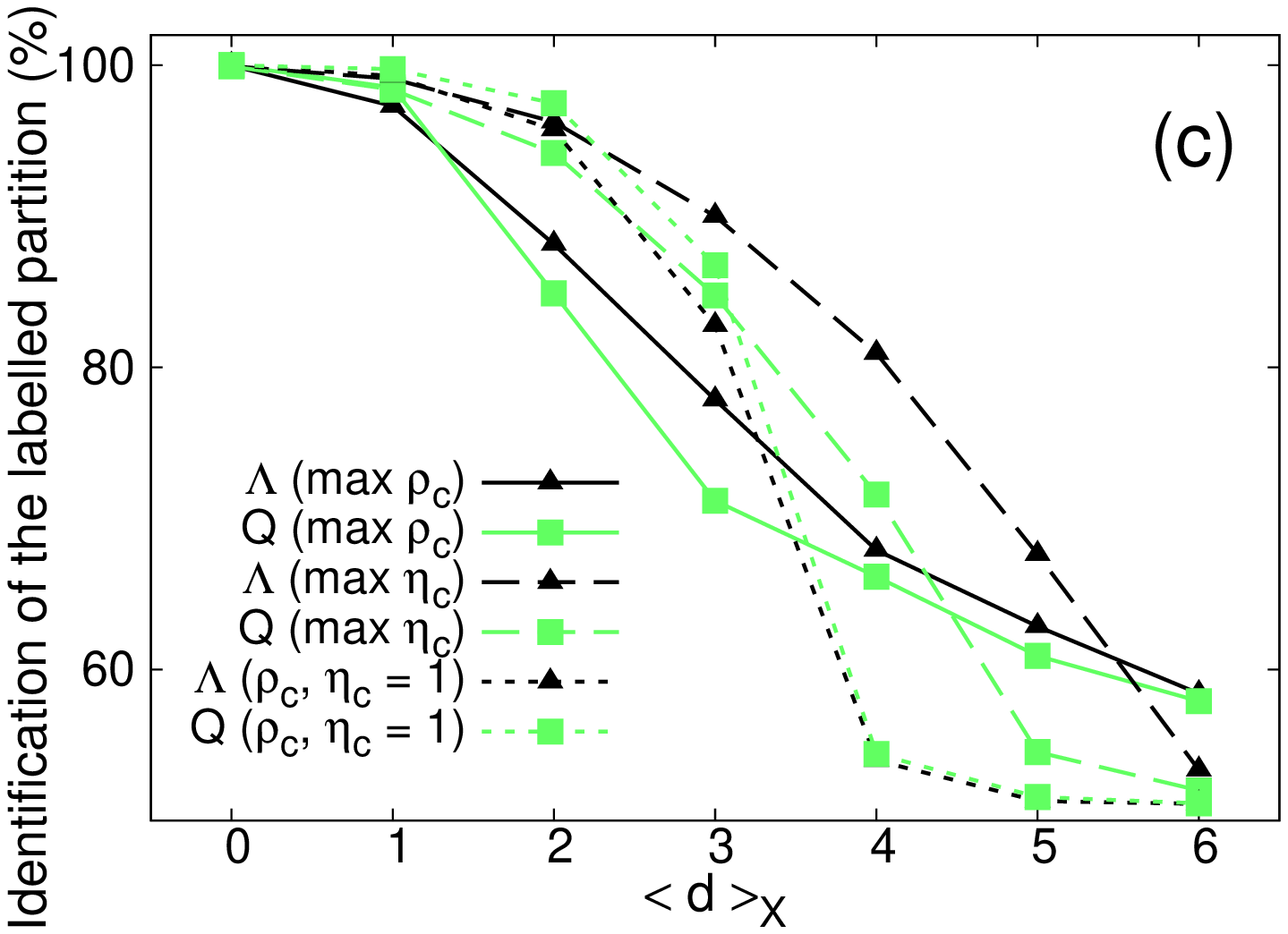}}
    \end{tabular}
    \caption{(Color online) Erdos-Renyi type networks with $N=1400$, $\langle d \rangle_I = 6$ and two communities.
    (a)  The average value of $\Lambda$ function (Eq. \ref{eq:spectral_cohesion}) for the partitions obtained by maximizing the spectral cohesion function and the modularity function.
    (b)  Average percent of nodes common between communities obtained by maximizing $\Lambda$ and $Q$.
    (c)  Average percent nodes of the labelled partition identified by $\Lambda$ and $Q$.
    Dark (black) colored curves are for the spectral cohesion function while the light (green) colored curves are for the modularity function.
    Data points represent averages over 20 simulated networks.}\label{fig:assort_n_corr_plots_er}
    \end{center}
\end{figure*}

\begin{figure*}
    \begin{center}
    \begin{tabular}{ccc}
        \resizebox{55mm}{!}{\includegraphics[width=3.75cm,height=5.00cm,angle=-90]{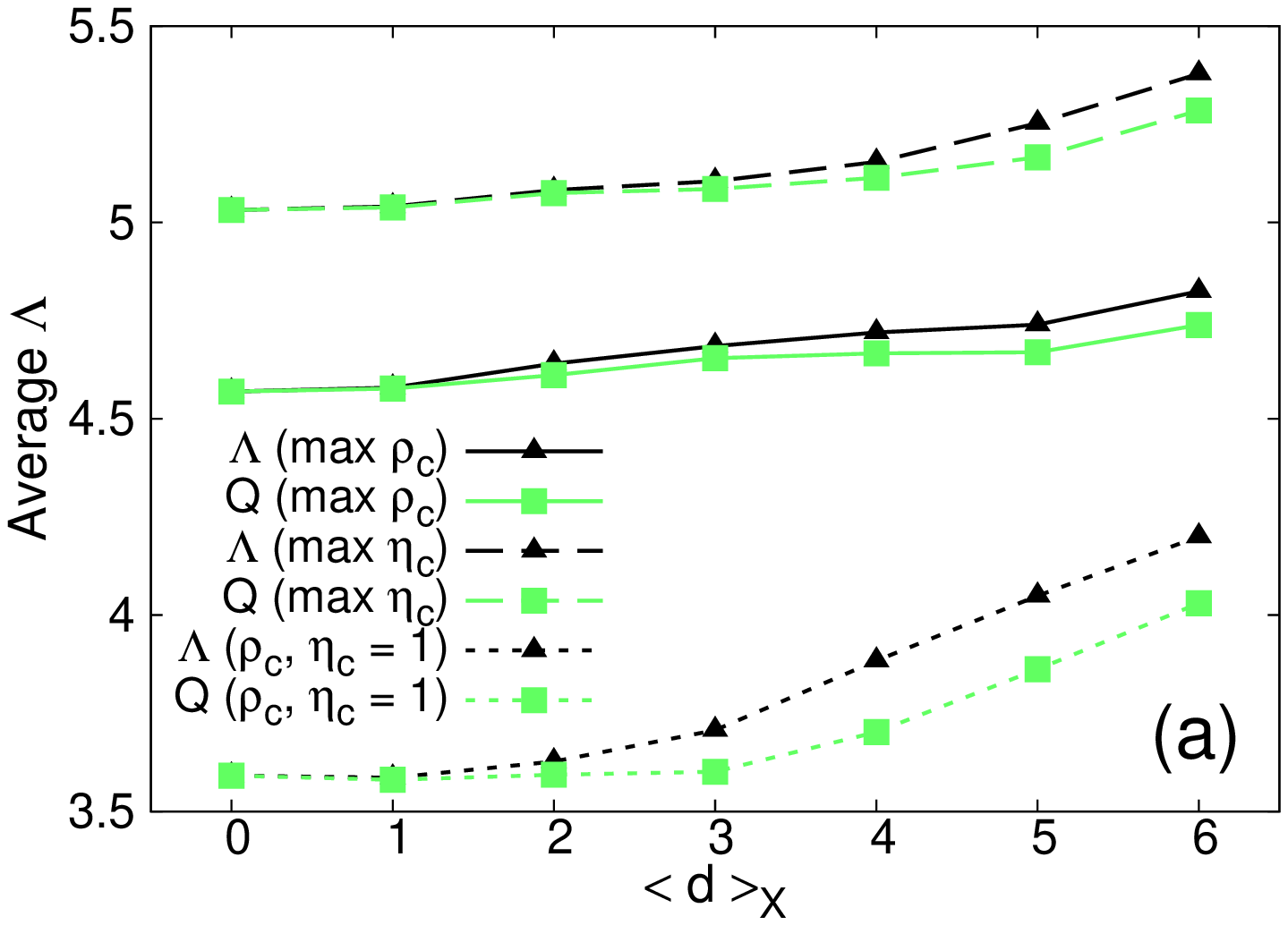}} &
        \resizebox{55mm}{!}{\includegraphics[width=3.75cm,height=5.00cm,angle=-90]{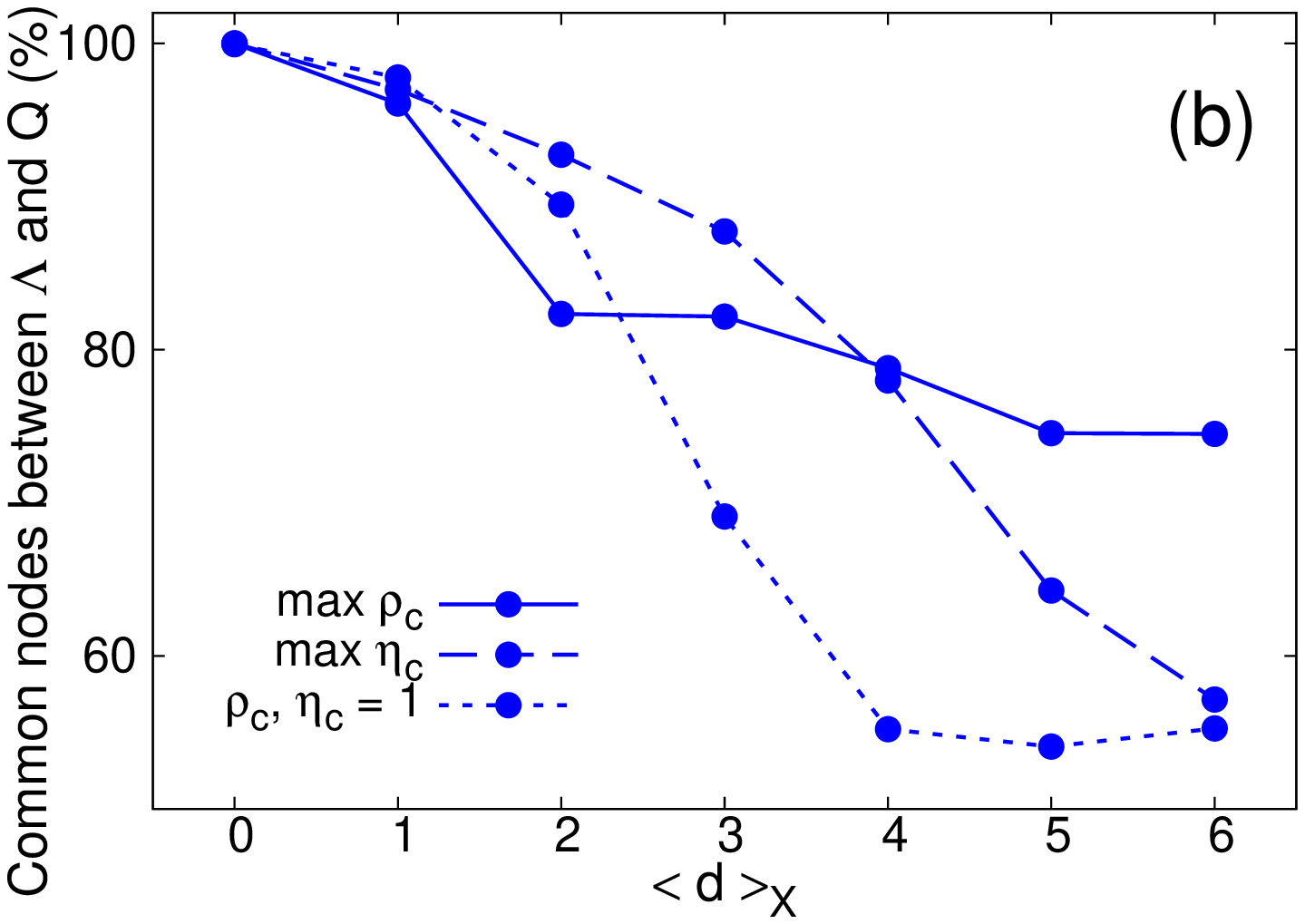}} &
        \resizebox{55mm}{!}{\includegraphics[width=3.75cm,height=5.00cm,angle=-90]{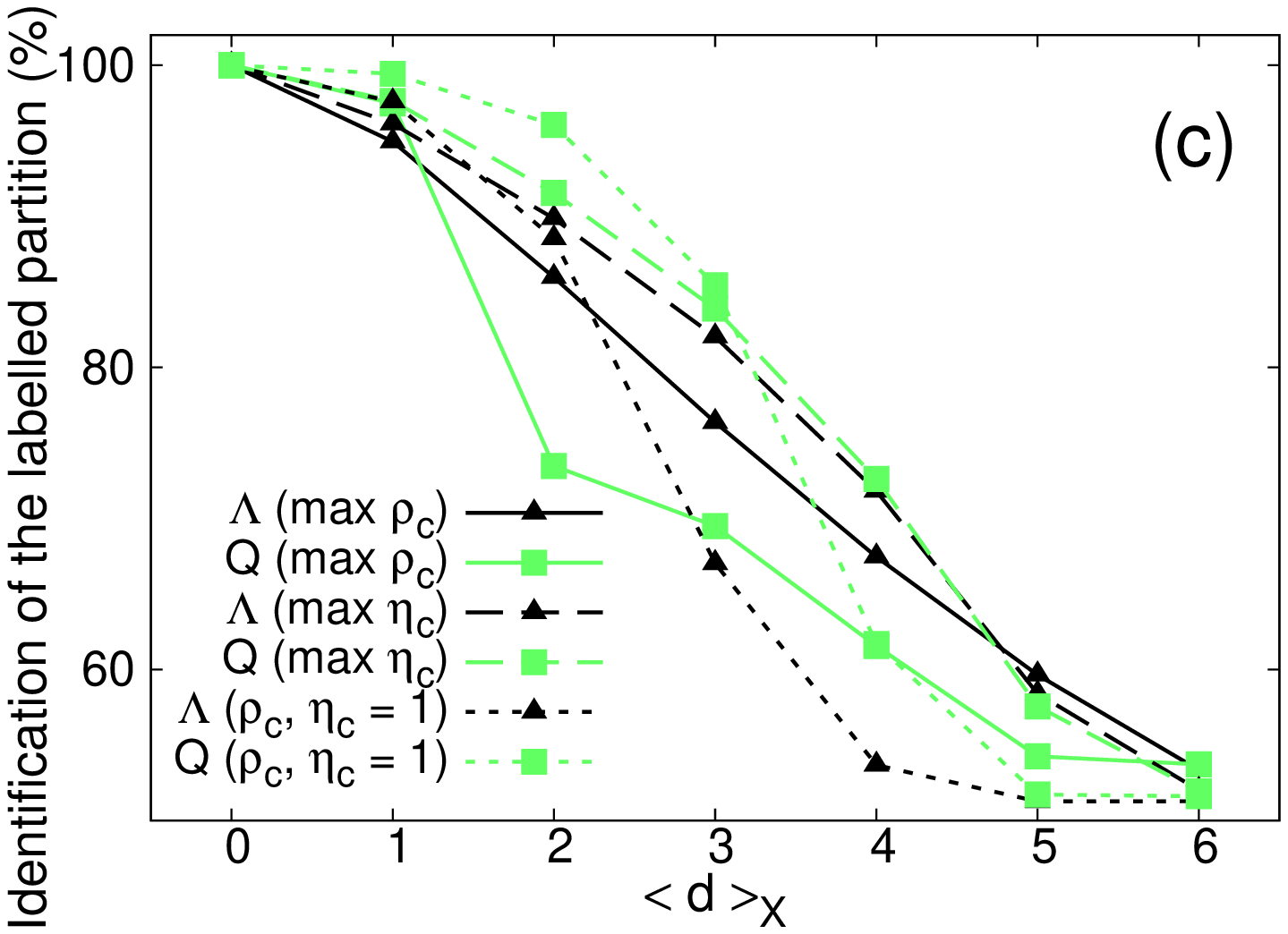}}
    \end{tabular}
    \caption{(Color online) Scale-free networks with $N=1400$, $\langle d \rangle_I = 6$ and two communities.
    (a)  The average value of $\Lambda$ function (Eq. \ref{eq:spectral_cohesion}) for the partitions obtained by maximizing the spectral cohesion function and the modularity function.
    (b)  Average percent of nodes common between communities obtained by maximizing $\Lambda$ and $Q$.
    (c)  Average percent nodes of the labelled partition identified by $\Lambda$ and $Q$.
    Dark (black) colored curves are for the spectral cohesion function while the light (green) colored curves are for the modularity function.
    Data points represent averages over 20 simulated networks.}\label{fig:assort_n_corr_plots_sf}
    \end{center}
\end{figure*}

Here we present results from applying our eigenvalue-based measure $\Lambda$ and modularity $Q$ to divisions of test networks into two communities via the simulated annealing procedure described in Section \ref{annealing}.
Results for Erdos-Renyi type networks are shown in Fig. \ref{fig:assort_n_corr_plots_er}, while results for scale-free networks are shown in Fig. \ref{fig:assort_n_corr_plots_sf}.  Each data point in Figs. \ref{fig:assort_n_corr_plots_er} and \ref{fig:assort_n_corr_plots_sf} represents an average over 20 random network realizations.

Figures \ref{fig:assort_n_corr_plots_er} (a) and \ref{fig:assort_n_corr_plots_sf} (a) show the spectral cohesion $\Lambda$ versus $\langle d \rangle_X$ for three different sets of network parameters [$\rho_c$ maximized with $\eta_c \approx 1$; $\eta_c$ maximized with $\rho_c \approx 1$; and ($\rho_c,\eta_c$) $\approx (1,1)$] when $\Lambda$ is maximized (plotted as solid triangles) and when $Q$ is maximized (plotted as solid squares).  Figures \ref{fig:assort_n_corr_plots_er} (b) and \ref{fig:assort_n_corr_plots_sf} (b) show the percent of nodes that are common between the $\Lambda$-based and the $Q$-based community partitions for the three network parameter sets.
Figures \ref{fig:assort_n_corr_plots_er} (c) and \ref{fig:assort_n_corr_plots_sf} (c) show the extent to which the $\Lambda$-based and the $Q$-based community divisions correspond to the ``labelled partition''.
By the labelled partition, we mean the partition with two equally sized communities into which we divide the nodes when we generate random networks.

Referring to Figs. \ref{fig:assort_n_corr_plots_er} (a) and \ref{fig:assort_n_corr_plots_sf} (a), we take the point of view that, essentially by definition, the $\Lambda$-based divisions give the best functional communities.  It is notable from these plots that, although $Q$-based divisions give lower than optimal $\Lambda$, the $Q$-division results for $\Lambda$ are surprisingly close to optimal throughout the whole range of $\langle d \rangle_X$ plotted.  In contrast, Figs. \ref{fig:assort_n_corr_plots_er} (b) and \ref{fig:assort_n_corr_plots_sf} (b) show that the percent agreement on nodal divisions between the $\Lambda$-based and the $Q$-based divisions can become substantial at large values of $\langle d \rangle_X$, especially for the networks with $\eta_c$ maximized and with ($\rho_c,\eta_c$)$\approx 1$, while agreement is significantly better for networks with $\rho_c$ maximized.

Regarding the difference between Figs. \ref{fig:assort_n_corr_plots_er} (a) and (c) and between Figs. \ref{fig:assort_n_corr_plots_sf} (a) and (c), we expect both community finding methods to yield imperfect identification of the labelled partition.  For example, this could result because it could happen that, in the random realization of a given network, some nodes with low within-community degrees in the labelled partitions may end up having many links with nodes in the other community or may get linked to high degree nodes in the other community.  In the test networks, such nodes would reasonably be classified as belonging to the community to which they were not originally assigned in the labelled partition.

\subsection{Networks with biased links between communities}
\label{biased_networks}

\begin{figure*}
    \begin{center}
    \begin{tabular}{cc}
        \resizebox{80mm}{!}{\includegraphics[width=3.0cm,height=4.0cm,angle=-90]{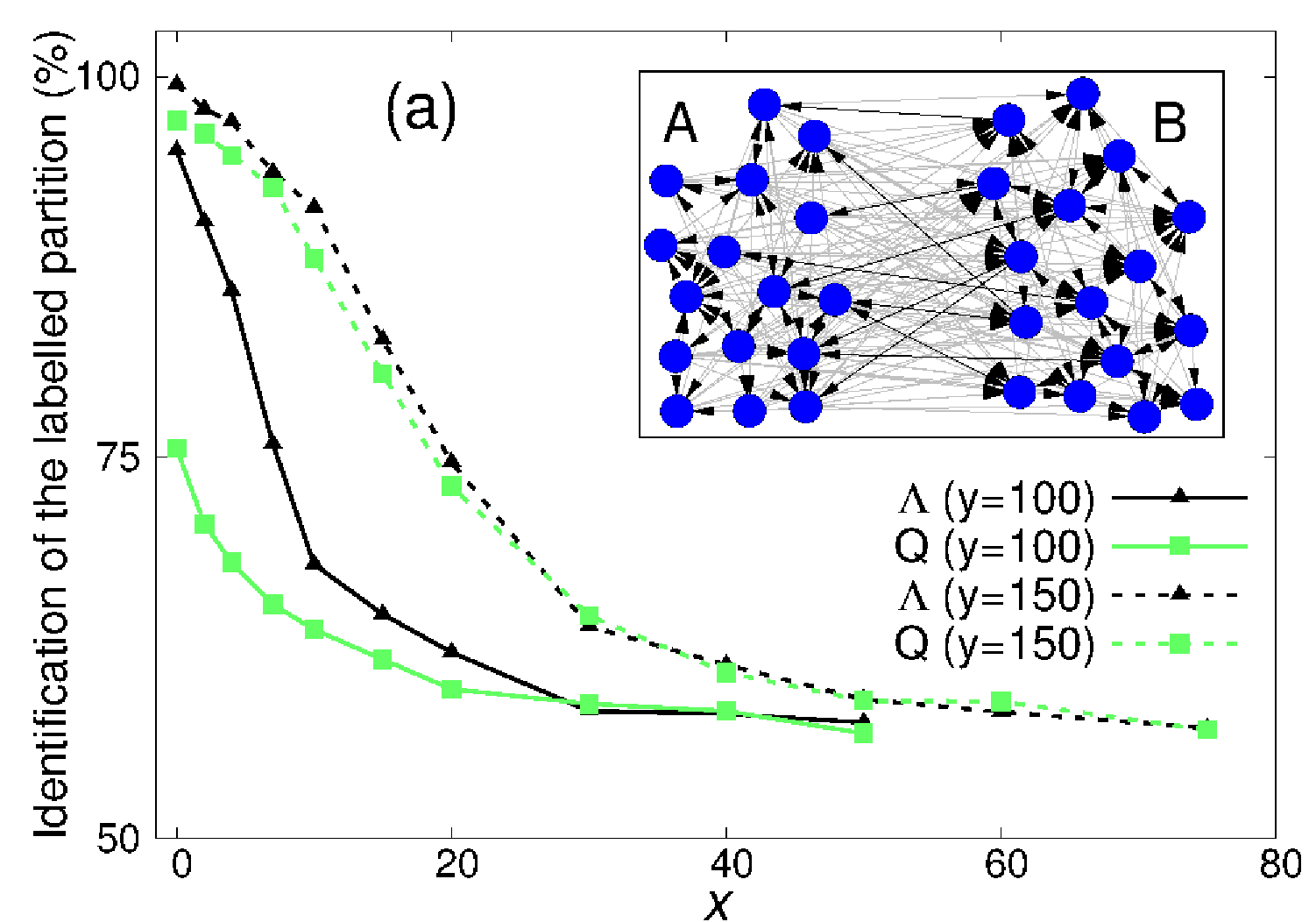}} &
        \resizebox{80mm}{!}{\includegraphics[width=3.0cm,height=4.0cm,angle=-90]{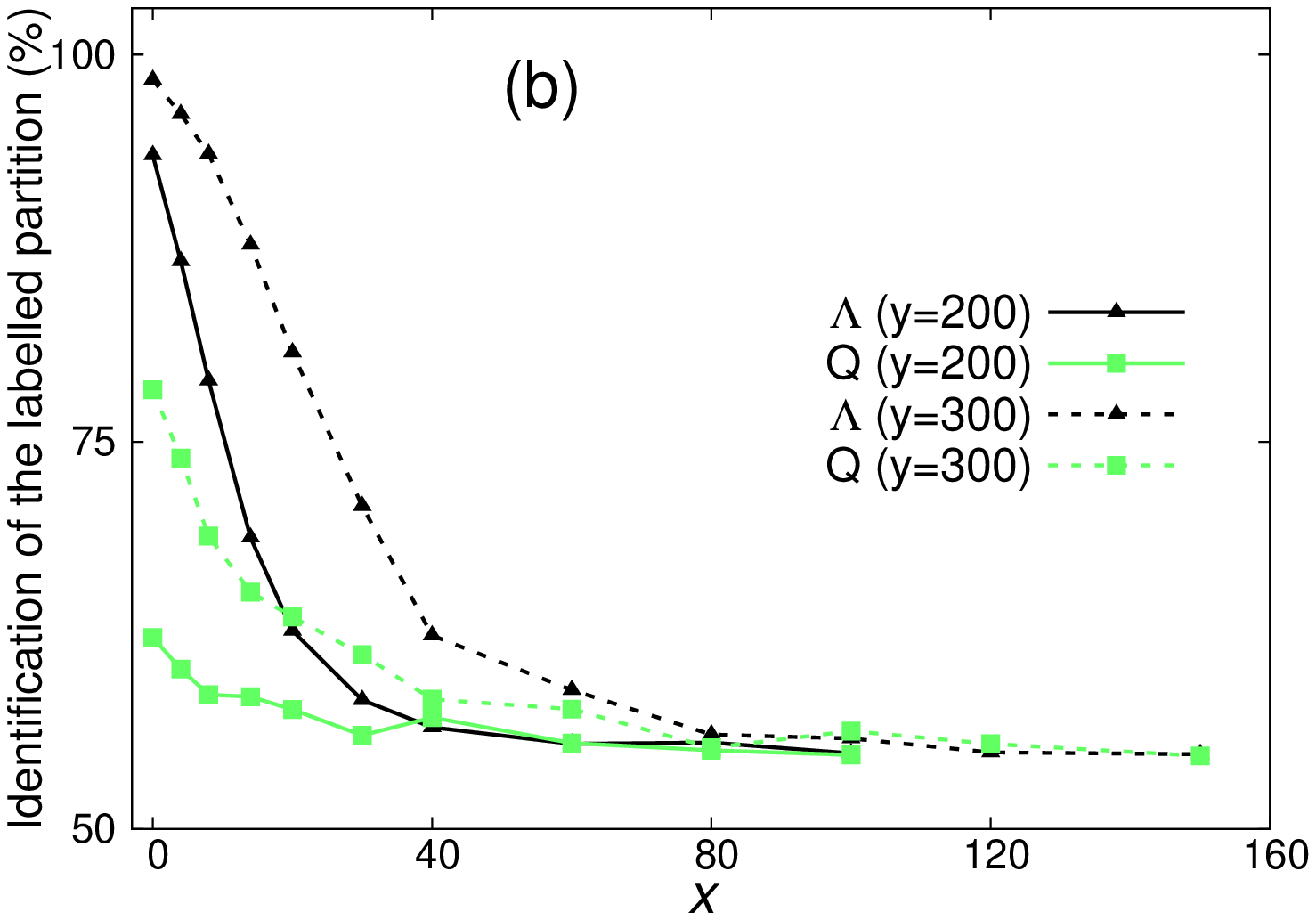}}
    \end{tabular}
    \caption{(Color online) Percent of the labelled partition identified by $\Lambda$ and $Q$ maximization vs `$x$' for the computer generated directed networks as explained in the text.  (a) Networks with 32 nodes.  (b) Networks with 64 nodes.  All data points are averaged over 100 network realizations.  The inset in figure (a) shows a particular network realization with $N=32$, $y=100$ and $x=10$.  For the sake of clarity, the 10 directed links that point from community $B$ to community $A$ are given darker shade.
    Note that, for the 32 node networks with $y=100$ and $x=0$ and the 64 node networks with $y=200$ and $x=0$, our $\Lambda$-based method does not give 100\% identification of the labelled partition.  This is because there are a few nodes in our random network realizations that are not part of the giant strongly connected component of either of the communities.  For these cases, as we increase the density of links in the networks ($y=150$ for the 32 node networks and $y=300$ for the 64 node networks), the probability of such nodes becomes much smaller and the identification rate for the $\Lambda$-based method at $x=0$ becomes close to 100\%.
    }\label{fig:dir_biased}
    \end{center}
\end{figure*}

Here, we consider directed networks with two communities of equal sizes.  We construct these networks so that, when the directionality of links is neglected, we get undirected random networks without communities.  To do this, we start with 32 nodes that are divided into two groups of equal sizes, where each group represent a community.  We then create, say, $y$ number of undirected links between the two groups of nodes and $y/2$ randomly oriented directed links within each group.  All the undirected links between the two groups are made directed with a bias such that more links point from one group of nodes to the other than the other way around.  Thus, when we have $x$ directed links pointing from one group to the other, $y-x$ directed links point in the opposite direction.  Varying $x$ gives us networks with a varying degree of community strength.  The results for these networks corresponding to $N=32$ and $N=64$ are shown in Fig.\ref{fig:dir_biased}.  At low values of $x$, when we have more bias, the spectral cohesion does better than modularity.  At relatively higher values of $x$, both functions give similar results.  We find that as we increase the number of links in the networks, by increasing the value of $y$, both the methods show improvement.

Comparing Fig. \ref{fig:dir_biased} (a) ($N$=32) and Fig. \ref{fig:dir_biased} (b) ($N$=64), we see that increasing the size of the networks keeping the average degree constant, the relative advantage of the $\Lambda$-based partitions as compared to the $Q$-based partitions increases substantially.

Thus, we see that when functional communities are very strongly dependent on link directionality, modularity may substantially under-perform compared to the spectral cohesion method.  We note, however, that so far we have not been successful at finding examples of real networks with this property.

\subsection{Discovering communities in real world networks}

To test how well our method finds communities in real networks, we used the networks of political blogs \cite{adamic2005} and jazz bands \cite{gleiser2003}.  The political blogs network is a directed network of weblogs on US politics during the 2004 US presidential elections.  The edges are the hyperlinks connecting two blogs.  The data for the network of jazz bands was obtained from The Red Hot Jazz Archive digital database.  The network consists of bands that performed between 1912 and 1940.  In this network, two bands are connected if there is a musician that played in both the bands.

The political blogs network has 1224 nodes with $\langle d \rangle =15.6$.  The eigenvalue plot of the adjacency matrix of this network shows two positive real eigenvalues well separated from the cloud of the rest of the eigenvalues \cite{paper1}.  This implies that there are two well defined communities in this network.  The communities apparently correspond to left/liberals and right/conservatives.  We used our simulated annealing procedure to divide the network into two communities by maximizing $\Lambda$ and $Q$.  Results are shown in Table \ref{tab:table2}, which also gives the percent of nodes common for the spectral cohesion method and the modularity method.  The values of $\Lambda$ for the spectral cohesion method and the modularity method are very close.  We believe that this is due to the fact that the two communities have giant strongly connected components that are well separated from each other.  Of the nodes that belonged to the giant strongly connected components of the network, there were 97.5 percent nodes common between the two community finding methods.
In this network, there were 431 nodes that did not belong to any community's strongly connected component.  Thus, in the spectral cohesion method, they were assigned based on the number of links such nodes had to the nodes in the giant strongly connected components of the communities.

\begin{figure}[tb]
\centering
\includegraphics[width=8.0cm,height=5.3cm,angle=0]{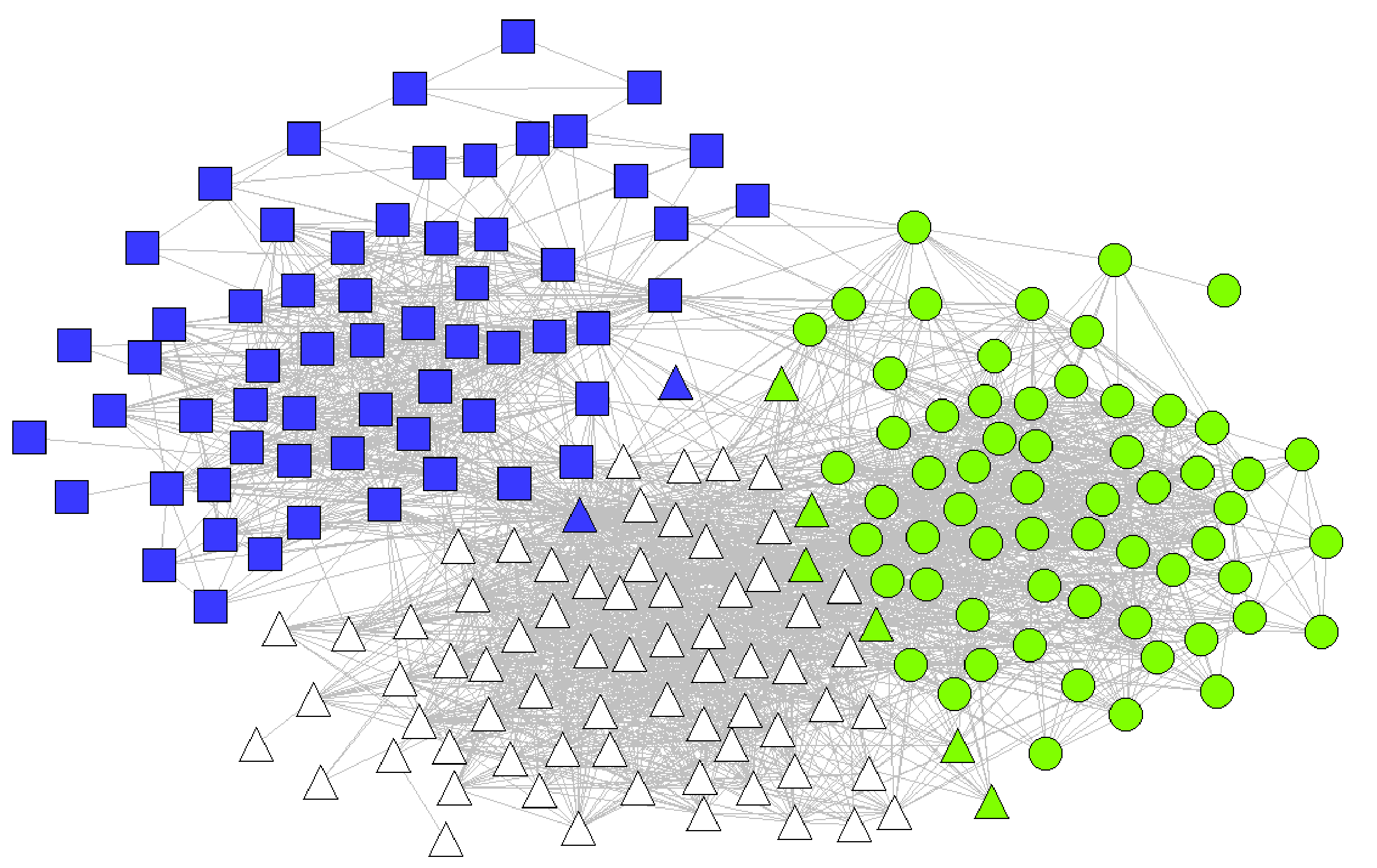}
\caption{(Color online) Comparison between the spectral cohesion method and the modularity method for the jazz bands networks.  Different shapes of the nodes correspond to communities obtained by maximizing $\Lambda$, while different colors correspond to communities obtained by maximizing $Q$.}\label{fig:jazz_bands}
\end{figure}

The network of jazz bands is an undirected network with 198 nodes and $\langle d \rangle=27.7$.  The eigenvalue plot of this network shows three positive eigenvalues that are well-separated from the bulk of the other eigenvalues, thus indicating three strong communities \cite{paper1}.  Two strong communities in this network correspond to predominantly the white bands and the black bands, which shows racial segregation.  The community of black bands divides further into two groups, the bands that performed in two major US cities, Chicago and New York \cite{gleiser2003}.  Figure \ref{fig:jazz_bands} shows the comparison between the partitions obtained by maximizing $\Lambda$ and $Q$.  We see that for this network, both the methods yield nearly the same network divisions (also see Table \ref{tab:table2}).

\begin{table*}[tb]
  \begin{center}
    \begin{tabular}{  c | c | c | c | c | c | } \cline{2-5}
        & \multicolumn{2}{|c|}{Optimizing $\Lambda$} & \multicolumn{2}{|c|}{Optimizing $Q$} \\ \hline 
        \multicolumn{1}{|c|}{Network}          & $\Lambda$  &    $Q$ &  $\Lambda$ &    $Q$  & \% common nodes \\ \hline
        \multicolumn{1}{|c|}{Political blogs}  &   6.817      &  0.416 &    6.816     &  0.431  & 94.7  \\ \hline
        \multicolumn{1}{|c|}{Jazz bands}  &   10.095      &  0.441 &    10.084     &  0.444  & 96.0  \\ \hline
    \end{tabular}
    \caption{Function values and percent nodes common for the real networks considered in this section.}
    \label{tab:table2}
  \end{center}
\end{table*}

\section{Conclusions}
\label{conclusions}

In this paper, we explored the utility of functional rather than structural definitions of community structure.  Specifically, as an example, we considered a definition of communities appropriate to cases where the communities are thought to form to enhance synchronizability and/or robustness to random node failures.  Our method is based upon our introduction of the spectral cohesion function $\Lambda$ (Eq. (\ref{eq:spectral_cohesion})) and is motivated by the role played by the maximum eigenvalue of the adjacency matrix, $\lambda_*$, in network functions.

Our study finds, perhaps, the unexpected result that for partitions obtained by maximizing modularity, the spectral cohesion, $\Lambda$, values are often close to optimal (Figs. \ref{fig:assort_n_corr_plots_er} (a) and \ref{fig:assort_n_corr_plots_sf} (a)) even when the modularity maximized partitions were substantially different from the $\Lambda$-maximized partitions (Fig. \ref{fig:assort_n_corr_plots_er} (b) and Fig. \ref{fig:assort_n_corr_plots_sf} (b)).  Although our eigenvalue based method is computationally intensive, our analysis shows that [except when communities are strongly dependent on link directionality (Sec. \ref{biased_networks})] communities obtained using the modularity-based method often do quite well when evaluated by our functional measure.

\appendix*

\section{Generating networks with the eigenvalue based communities}
\label{generating_nets}

\subsection{Scale-free networks with nodal in/out-degree correlation within communities}
\label{sf_node_corr}

To generate these networks, we start by dividing the nodes into two equally sized communities (labelled by $k=1,2$).  For
nodes in community $k$, we then generate two degree sequences corresponding to
the in-degrees ($d_i^{in}$) and the out-degrees ($d_i^{out}$),
according to the power law degree distribution, $P(d)\propto d^{-\gamma}$, by using the formula \cite{restrepo2006}:
\begin{equation} \label{eq:sf_numbers}
b(l+l_0-1)^{-1/(\gamma-1)}
\end{equation}
for $l=1,2,3,...,N_k$.  Here, the constants $b$ and $i_0$ determine the maximum degree and node averaged degree.  For the test networks used in this paper, we used $\gamma=2.5$.
These $N_k$ numbers corresponding to both the degree sequences are assigned randomly to the $N_k$ nodes in community $k$.  Since these numbers are assigned independently at random, $d_i^{in}$ and $d_i^{out}$ of the nodes are uncorrelated.  We perform this procedure for each of the communities, separately.

For each of the $d_i^{in}$ for node $i$, we then randomly divide $d_i^{in}$ into two compartments,
\begin{equation} \label{eq:in-degree_decomposition}
d_i^{in} = (d_i^{in})_I + (d_i^{in})_X,
\end{equation}
choosing $0 \le (d_i^{in})_I \le d_i^{in}$ from a binomial distribution.  Here the subscripts $I$ and $X$ signify internal and external, and $(d_i^{in})_I$ signifies the number of links going to node $i$ from nodes in its own community, while $(d_i^{in})_X$ signifies the number of links going to node $i$ from nodes that are not in its community.  In addition, we perform a similar decomposition for $d_i^{out}$,
\begin{equation} \label{eq:out-degree_decomposition}
d_i^{out} = (d_i^{out})_I + (d_i^{out})_X.
\end{equation}
In using the binomial distribution in Eq. (\ref{eq:in-degree_decomposition}) (or Eq. (\ref{eq:out-degree_decomposition})), we assume that each of the $d_i^{in}$ links (or $d_i^{out}$ links) has probability $\langle d \rangle_I/\langle d \rangle$ of being internal.  We now have associated to each node $i$ the four degrees
\begin{equation} \label{eq:four_degrees}
[(d_i^{in})_I, (d_i^{in})_X, (d_i^{out})_I, (d_i^{out})_X].
\end{equation}
To create a network with maximal $\eta_c$, we now shuffle the node assignments of $[(d_i^{in})_I,(d_i^{in})_X]$ within each community, while keeping the node assignments of $[(d_i^{out})_I,(d_i^{out})_X]$ fixed.  In particular, if $(d_j^{out})_I$ is the largest internal out-degree of community $k$, we reassign $[(d_i^{in})_I,(d_i^{in})_X]$ from node $i$ to node $j$ where node $i$ is also the node in community $k$ with the largest value of $(d_i^{in})_I$.  We then do the same for the second largest, for the third largest, etc.  This leads to new assignments of the four degree quantities (Eq. (\ref{eq:four_degrees})) for each node but only by shuffling degrees of nodes that belong to the same community.  Note than, by construction, our reassignment procedure leaves the distribution of the $d_i^{in}$ and $d_i^{out}$ invariant, and that interchanging the roles of ``in'' and ``out'' (\emph{i.e.}, preserving the ``in'' node assignments and shuffling the ``out'' node assignments) results in an equivalent procedure.

We now construct within community links for each community $k$.  We imagine drawing $(d_i^{in})_I$ in-stubs and $(d_i^{out})_I$ out-stubs at each node $i$.  We then randomly pair the end of an in-stub to the end of an out-stub and connect them with a link.  This is done avoiding repeated links and self-links.  Finally, we use the analogous procedure to construct external connections between communities.

When we generate scale-free networks, nodes with high within community in-degrees (out-degrees) will tend to have high between community in-degree (out-degree).  Similarly, nodes with low within community degrees will tend to have low between community degrees.  We do this with the belief that important nodes that have high number of links within their own community will, in general, have high number of links attached to nodes outside their own community.  Due to this, when we generate scale-free networks with maximal node degree correlations within the communities, they also tend to have higher values of node degree correlations for the between community degrees, although for each node, we expect the between community in and out-degrees to be less correlated than the within community in and out-degrees.

When we generate scale-free networks with the maximal node degree correlations within the communities, we find that the values of the directional degree assortativity within the communities become slightly less than 1.  Since we are interested in looking at the effect of changing nodal degree correlations within the communities, we use the edge swapping procedure described in Sec. \ref{edge_degree_corr} to restore $\rho_c$ to $\rho_c\approx 1$ within the communities.

\subsection{Scale-free networks with directional degree assortativity within communities}
\label{sf_assorted}

We construct a directed random node degree uncorrelated
scale-free network by using the method given in Appendix \ref{sf_node_corr}
(\emph{i.e.}, without shuffling the degrees of the nodes).
We then use the edge swapping procedure given in Sec. \ref{edge_degree_corr} to get maximal possible $\rho_c$ within each community by considering only within community degrees, $(d^{in}_i)_I$ and $(d^{out}_i)_I$.

\subsection{Erdos-Renyi type networks with nodal in/out-degree correlation within communities}
\label{er_node_corr}

To get these networks, we first divide the nodes into two equally sized communities and create undirected edges within communities with probability, say $p$.  Considering undirected edges as bidirected links, we randomly reassign links between nodes keeping their degrees constant.  This gives us communities in which the in-degree of a node equals the out-degree of the node but the edge degree correlations are absent.  Between community directed links are created by creating directed links between pairs of nodes in different communities with some other chosen probability, say $q$.

\subsection{Erdos-Renyi type networks with directional degree assortativity within communities}
\label{er_assorted}

To generate these networks, we divide the nodes in the network into two equally sized communities.  Within communities, we create directed links with probability $p$ while between community directed links are created with some other probability $q$.  We then use the procedure of Sec. \ref{edge_degree_corr} to get maximal $\rho_c$ within the communities.

\begin{acknowledgments}
This work was supported by ONR grant N00014-07-1-0734.
\end{acknowledgments}


\bibliography{paper}		

\end{document}